\documentstyle[prd,preprint,aps,floats]{revtex}

\begin{document}
%%%%%%%%%%%%%%%%%%%%%
%title page
%%%%%%%%%%%%%%%%%%%%%

%%%%%%%%%%%%%my-commands%%%%%%%%%%%%%%%%%%%%%%%%%%%%%%%%%%%%%%%
\newcommand{\be}{\begin{equation}}
\newcommand{\ee}{\end{equation}}
\newcommand{\bea}{\begin{eqnarray}}
\newcommand{\beas}{\begin{eqnarray*}}
\newcommand{\eea}{\end{eqnarray}}
\newcommand{\eeas}{\end{eqnarray*}} 
\newcommand{\ba}{\begin{array}}
\newcommand{\ea}{\end{array}}
%%\renewcommand{\theequation}{\arabic{section}.\arabic{equation}}
%%%%%%%%%%%%%newmath-commands%%%%%%%%%%%%%%%%%%%%%%%%%%%%%%%%%%%%%%%
\def\ls{\mathrel{\lower4pt\vbox{\lineskip=0pt\baselineskip=0pt
           \hbox{$<$}\hbox{$\sim$}}}}
\def\gs{\mathrel{\lower4pt\vbox{\lineskip=0pt\baselineskip=0pt
           \hbox{$>$}\hbox{$\sim$}}}}
%%%%%%%%%%%%%%%%%%%%%%%%%%%%%%%%%%%%%%%%%%%%%%%%%%%%%%%%%%%%%%%%

\tightenlines

\def\DESepsf(#1 width #2){\epsfxsize=#2 \epsfbox{#1}}
\draft
%\begin{titlepage}
\thispagestyle{empty}
\preprint{\vbox{ \hbox{UMD-PP-02-55}
\hbox{May, 2002}}}

\title{ \Large\bf Left-Right Symmetry in 5-D and Neutrino Mass in
TeV-scale Gravity Models}

\author{\large
R.N. Mohapatra$^1$\footnote{e-mail:rmohapat@physics.umd.edu},
and A. P\'erez-Lorenzana$^{2,3}$\footnote{aplorenz@ictp.trieste.it}}

\address{$^1$Department of Physics, University of Maryland\\
College Park, MD 20742, USA\\
$^2$The Abdus Salam International Centre for Theoretical Physics,
I-34100, Trieste, Italy\\
$^3$Departamento de F\'\i sica,
Centro de Investigaci\'on y de Estudios Avanzados del I.P.N.\\
Apdo. Post. 14-740, 07000, M\'exico, D.F., M\'exico}

\maketitle

\thispagestyle{empty}

\begin{abstract}
We construct a left-right symmetric model based on the gauge group
$SU(2)_L\times SU(2)_R\times U(1)_{B-L}$ in five dimensions where
both the gauge bosons and fermions reside in all five
dimensions. The orbifold boundary conditions are used not only to break
the gauge symmetry down to $SU(2)_L\times U(1)_Y\times U(1)_{Y'}$ but
 also to ``project''  the right handed neutrino out of the zero
mode part of the spectrum, providing a new way to understand the small
neutrino masses without
adding (singlet) bulk neutrinos. This formulation of the left-right model
has also two new features: (i) it avoids most existing phenomenological
bounds on the scale of the right handed $W_R$ boson allowing for the
possibility that the right handed gauge bosons could have masses under a
TeV, and (ii) it predicts a stable lepton with mass of order of
the inverse radius of the fifth dimension.
 \end{abstract}

%%%%%%%%%%%%%%%%%%%%%%%%%%%%%
\section{Introduction}

 Left-right symmetric models of weak interactions were introduced 
to understand the origin of parity violation in weak
interactions~\cite{moh}. Several features that have made this class of
models interesting are: (i) complete quark-lepton symmetry, (ii) a more
physically meaningful formula for the
electric charge than the standard model~\cite{marshak}, (iii) a natural
way to understand small neutrino masses via the seesaw
mechanism~\cite{seesaw}, (iv) solution to the strong CP
problem~\cite{goran}, (v) suppression of R-parity violating operators
present in the minimal extension of the standard model due
to local $B-L$ symmetry\cite{rp}; (vi) phenomenology of
the extra $Z'$ boson\cite{extra}. Many recent string
constructions have also
led to left-right symmetric models below the string scale\cite{ibanez}.

An important question for phenomenology of these models has been the scale
of parity
breaking. For a long time there has been interest in models where the
masses of the right handed $W_R$ and $Z'$ are in the multi-TeV range.
 With the recent discovery
of nonzero neutrino masses, the case for left-right models has become more
compelling for two reasons: (i) the right handed neutrino, which is
necessary to
implement the seesaw mechanism, is an integral part of these
models and (ii)
the local $B-L$ symmetry which protects the right handed neutrino mass
from being at the Planck scale is also part of the gauge
symmetry. However,
the present observations of neutrino oscillations, coupled with bounds
on neutrino masses from tritium beta decay require that if the
seesaw mechanism is to explain neutrino masses, then the
scale of parity breaking
must be very high (of order $10^{12}$ GeV or higher). This makes the
effects associated with right handed gauge symmetry (such as the
extra $Z'$ boson, right handed current
induced flavor changing effects, right handed
current effects in weak decays etc.) completely unobservable. An
interesting question therefore is whether it is possible to have a low
scale for $SU(2)_R$ symmetry and still have a natural understanding of
small neutrino masses with a minimal particle content.
With the recent revival of theories with low scale for new
physics in the context of brane bulk picture of space time, this question
becomes a very pertinent one. It is quite
reasonable to ask whether a case can be made for the right handed scale
being in the multi-TeV range close to the fundamental scale of nature and
yet have a solution to the neutrino mass
problem. Coupled with the fact that the brane-bulk models
 provide an attractive alternative to supersymmetry as a way to
solve the gauge hierarchy problem~\cite{arkani}, one would now have 
an additional reason to consider the multi-dimensional low scale models
as the ultimate theory of nature below the string scale.

The most widely discussed scenarios of the brane-bulk type put the
standard model
fermions in the brane and gauge bosons either in the brane or the
bulk. Either of these cases have the difficulty that they cannot
accommodate the attractive seesaw mechanism for the neutrino
masses. As a result, to solve the neutrino mass problem 
one adds two new ingredients:
 (i) extra singlet fermions\cite{nima2} in the bulk to suppress the
Dirac mass of the neutrino and (ii) a global symmetry such as $B-L$ to
prevent dangerous higher dimensional operators
of the form $(LH)^2/M_*$ ($L,H$ are  lepton and Higgs doublets of the
standard model respectively). On the one hand, the bulk neutrinos can
cause drastic
revisions of our understanding of such issues as big bang nucleosynthesis
and supernova dynamics and on the other hand, the assumption of global
$B-L$ symmetry contradict
the fact that string theories are not supposed to
have any continuous global symmetries\cite{witten}.

A natural question to ask is whether the above problems arise due to our
insistence that the gauge group is the standard $SU(2)_L\times U(1)_Y$
all the way upto the string scale. We examine this question 
by introducing the left-right
symmetric gauge structure in the low fundamental scale models
and see whether it can provide a simpler way of understanding the small
neutrino
masses without introducing the bulk neutrinos and without assuming global
$B-L$ symmetry. The additional advantage is that due to low $SU(2)_R$ 
scale, the right handed current induced
effects could possibly be accessible to experiments providing new
tests of low scale gravity models. In the present paper, we
present such a model in the five dimensional context.

Before proceeding to the presentation of our formulation of the five
dimensional left-right symmetry, we note that
in a recent paper~\cite{nandi}, Mimura and Nandi have constructed a brane
model with the left-right symmetric structure in 5-dimensional space time
where
the gauge bosons reside in all five dimensions and the fermions
are confined to the 3+1 dimensional brane. They showed that the orbifold
projections can break the gauge symmetry down to $SU(2)_L\times
U(1)_Y\times U(1)_{Y'}$. Understanding small neutrino masses in
this formulation would require new
ingredients since within the minimal Higgs sector adopted in
Ref.\cite{nandi}, one expects a large ($\sim$ weak scale) Dirac mass for
the neutrino.

In this paper, we present a new formulation of left-right symmetry in the
five dimensions, where
we allow both gauge bosons and fermions to reside in five
dimensions and use orbifold boundary conditions in such a way that they 
not only make the right handed charged
$W_R$ boson massive but also ``project out'' the right handed neutrino
from the standard model brane, 
and so, out of the zero mode part of the spectrum. 
This assumption leads to a number of
interesting consequences.

 (i) The  first implication of this new construction
is that there
is now no Dirac mass for the left handed neutrino; secondly there are no
 lower order nonrenormalizable operators that
could give a large mass to neutrinos; the lowest order operator in five
dimensions that contributes to neutrino masses is $D=10$; this allows us
to have a solution to the neutrino mass problem even though the
scale of right handed symmetry is only in the multi-TeV range. 
No (singlet) bulk neutrino need be invoked~\cite{nima2}.

(ii) Another consequence of our orbifold breaking is that many of the
conventional phenomenological constraints on the $W_R$ mass do not
apply; as for example the $W_L-W_R$ box graph\cite{soni} that provided a
stringent constraint on $W_R$ in usual left-right models is now
absent. Similarly, the collider constraints\cite{cdfd0} as well as
the low energy muon\cite{strovnik} and beta decay constraints\cite{beg}
are now forbidden by the five dimensional momentum conservation.

(iii) The model predicts the existence of a heavy stable particle 
(a lepton) whose mass
is equal to the inverse radius of the fifth dimension. This can lead to
interesting cosmological as well as phenomenological possibilities.

This paper is organized as follows: in sec. II, we discuss the orbifold
breaking of the gauge symmetry as well as the 
boundary conditions satisfied by the matter fields 
of the theory; in section III, we
discuss the masses and mixings of the gauge bosons; in section IV, we
present the gauge boson fermion couplings; in section V, we discuss
neutrino masses and the prediction of a heavy sterile neutrino; in section
VI, we present some phenomenological implications such as the
$K^0-\bar{K}^0$ mixing, decay modes of the $W_R$ etc. In section VII, we
speculate on further extensions of the model and inclusion of
supersymmetry. There we also discuss how parity restoration occurs on the theory.
In section VIII, we present a summary of our results
and conclude. 

 %%%%%%%%%%%%%%%%%%%%%%%%%%%%%%%
\section{Fermions, Higgs bosons and orbifold breaking of $SU(2)_R$}

We now discuss the detailed particle content of the model and the orbifold
conditions to implement symmetry breaking. We denote the gauge symmetry in
five dimensions as $SU(2)_1 \times SU(2)_2 \times U(1)_{B-L}$, with
the later identification of subscripts as 
$1\rightarrow L$, and $2\rightarrow R$. 
The gauge
bosons are denoted by $W^{\pm,3}_{1,M}$, $W^{\pm,3}_{2,M}$ and $B_M$, with
$M=\mu, 5$ ($\mu$ denotes the 3+1 Minkowski indices and $5$ corresponds to
the compactified fifth dimension). We choose four sets of quark
doublets denoted by $Q_1$, $Q'_1$, $Q_2$ and $Q'_2$ and similarly four
sets of
lepton doublets denoted by $\psi_{1}$, $\psi'_1$,  $\psi_2$ and
$\psi'_2$. The
subscripts (1,2) next to fermions represent that they transform as
doublets under the corresponding $SU(2)$ group. 
Notice that all fermions  contain both
left and right handed components, and five dimensional  interactions are
originally vectorlike.
Nevertheless, the non trivial boundary conditions introduced 
by orbifolding can  break  both, 
the gauge symmetry~\cite{scherk,kawamura} and the vectorlike
nature of fermions. We will use this feature
in what follows.

Let us now proceed with the orbifold compactification of the fifth
dimension. We compactify the fifth coordinate (denoted by $y$) on an
orbifold $S_1/Z_2\times Z'_2$, where under the first $Z_2$: 
$y\rightarrow -y$; and under the second $Z'_2$: $y'\rightarrow -y'$; where
$y'= y + \pi R/2$.  The orbifold is then an interval $[0, \pi R/2]$. 
Since both gauge
bosons and fermions in our model propagate in all five dimensions, we have
to specify their transformation properties 
under the $Z_2\times Z'_2$. 
This is equivalent to define the boundary conditions that those fields
should satisfy.

As far as the gauge fields go, we choose the same boundary conditions as
in Ref.~\cite{nandi}. Defining the gauge boson matrix for both the
$SU(2)$'s
as 
\begin{eqnarray}
W~=~ \left(\begin{array}{cc} W_3 & \sqrt{2} W^+\\ \sqrt{2} W^- & -W_3
\end{array} \right)
\end{eqnarray}
we can
write the $Z_2\times Z'_2$ transformation properties of $W_{1,2}$ as
\begin{eqnarray}
W_{\mu}(x^{\mu}, y)&\rightarrow& W_{\mu}(x^{\mu}, -y)
       ~~=~\quad PW_{\mu}(x^{\mu},y) P^{-1} \nonumber \\
W_{5}(x^{\mu}, y)&\rightarrow& W_{5}(x^{\mu}, -y)
       ~~=~-PW_{5}(x^{\mu},y) P^{-1} \nonumber \\  
W_{\mu}(x^{\mu}, y')&\rightarrow& W_{\mu}(x^{\mu},-y')
       ~=~\quad P'W_{\mu}(x^{\mu},y') P^{'-1} \nonumber \\
W_{5}(x^{\mu}, y')&\rightarrow& W_{5}(x^{\mu}, -y')
       ~=~-P'W_{5}(x^{\mu},y)  P^{'-1}
\label{wtrans}
\end{eqnarray}
where $P$, and $P'$ are two by two diagonal matrices that we chose as
(i) $P~=~P'~=~diag(1,1)$ for the $SU(2)_1$ gauge bosons; and 
(ii) $P~=~diag(1,1)$ and  $P'~=~ diag(1,-1)$, for those of $SU(2)_2$. 
The $B$ boson, on the other hand, has a single 
transformation property under both $Z_2$ and $Z_2'$ projections, which is:
$B_\mu(x^\mu,y)  \rightarrow B_\mu(x^\mu,-y)$ and 
$B_5(x^\mu,y)  \rightarrow  - B_5(x^\mu,-y)$; and same under $Z_2'$. 

Under these transformations the gauge fields  develop explicit 5th
dimensional parities identified as the $Z_2\times Z_2'$ quantum numbers 
(this is to be distinguished from the usual space time parity).
With above transformation rules we find the following parity
assignments: 
\begin{eqnarray}
&&W_{1,\mu}^{3,\pm}(+,+);\quad B_\mu(+,+);\quad W^3_{2,\mu}(+,+);
 \quad W^\pm_{2,\mu}(+,-); \nonumber\\
&&W_{1,5}^{3,\pm}(-,-);\quad B_5(-,-);\quad W^3_{2,5}(-,-);
 \quad W^\pm_{2,5}(-,+).
\label{gparity}
\end{eqnarray}
To be more explicit about the meaning of these parities, note that a minus
sign indicates that such an odd
field vanishes at the fixed point associated to the corresponding $Z_2$ or
$Z_2'$ transformation. Thus, from Eq.~(\ref{gparity}) we see that at  
$y=\pi R/2$ the $SU(2)_2$ gauge symmetry has been broken down to its
diagonal $U(1)_{I_{3,2}}$ subgroup, 
while the other groups remain unbroken.
Thus, at this fixed point the remaining symmetry can be  identified as
$SU(2)_1\times U(1)_Y \times U(1)_{Y'}$. 
That is the Standard Model (SM) symmetry
with an extra $U(1)_{Y'}$ factor generated by the orthogonal generator to
the hypercharge: 
${1\over 2 }Y'\equiv\sqrt{\frac{2}{5}} I_{3,2} -\frac{3}{\sqrt{10}} 
{1\over2} (B-L)$.
At the other boundary ($y=0$) the whole gauge symmetry remains intact.
Due to the breaking of the symmetry at one of the boundaries,
the effective four dimensional theory will be invariant only under 
$SU(2)_1\times U(1)_Y \times U(1)_{Y'}$ symmetry

Turning now to the fermions, the general transformation rules under
$Z_2\times Z_2'$ of any of our doublet 
fermion representations have the form
\bea
\Psi(x^\mu,y) &\rightarrow & \Psi(x^\mu,-y)~=~
  \pm\gamma_5 P \Psi(x^\mu,y);  \nonumber \\
\Psi(x^\mu,y') &\rightarrow & \Psi(x^\mu,-y')~=~
  \pm\gamma_5 P'\Psi(x^\mu,y');
\eea
where the sign controls which chiral component 
of $\Psi$ is being projected
out of the fixed points by the action of $\gamma_5$. 
In last  equation $P$ and $P'$ 
act on the group space and are given by 
the very same matrices used in Eq.~(\ref{wtrans}).
In conclusion, left and right
components of the same fermion will   hold  opposite parities.
Up and down components of any
doublet representations under $SU(2)_1$ will  hold same
parity assignments. 
For $SU(2)_2$ doublets the situation is as follows. 
Parity assignments under $Z_2$ for the up and down fields will be the same,
however, those associated to $Z_2'$ will be opposite to each other, due to
the non trivial election of $P'$ acting on this group sector.

Using these rules we demand that the various fermion doublets
get the following $Z_2\times Z_2'$ quantum numbers, for quarks:
 \bea
 Q_{1,L}\equiv 
   \left(\begin{array}{c} u_{1L}(+,+)\\ d_{1L}(+,+)\end{array}\right);
 &\quad& 
 Q'_{1,L}\equiv 
   \left(\begin{array}{c} u'_{1L}(+,-)\\ d'_{1L}(+,-)\end{array}\right);
   \nonumber \\ 
 Q_{1,R}\equiv
   \left(\begin{array}{c} u_{1R}(-,-)\\ d_{1R}(-,-)\end{array}\right);  
 &\quad& 
 Q'_{1,R}\equiv 
   \left(\begin{array}{c} u'_{1R}(-,+)\\ d'_{1R}(-,+)\end{array}\right);
   \nonumber \\ [1ex]
 Q_{2,L}\equiv 
   \left(\begin{array}{c} u_{2L}(-,-)\\ d_{2L}(-,+)\end{array}\right);
 &\quad& 
 Q'_{2,L}\equiv 
   \left(\begin{array}{c} u'_{2L}(-,+)\\ d'_{2L}(-,-)\end{array}\right); 
   \nonumber \\ 
 Q_{2,R}\equiv 
   \left(\begin{array}{c} u_{2R}(+,+)\\ d_{2R}(+,-)\end{array}\right);
& \quad& 
Q'_{2,R}\equiv 
   \left(\begin{array}{c} u'_{2R}(+,-)\\ d'_{2R}(+,+)\end{array}\right);
\label{quarks}
\eea
and for leptons: 
 \bea
 \psi_{1,L}\equiv 
   \left(\begin{array}{c} \nu_{1L}(+,+)  \\ e_{1L}(+,+)\end{array}\right); 
  &\qquad&
 \psi'_{1,L}\equiv 
   \left(\begin{array}{c} \nu'_{1L}(-,+)  \\ e'_{1L}(-,+)\end{array}\right); 
   \nonumber \\ 
 \psi_{1,R}\equiv 
    \left(\begin{array}{c} \nu_{1R}(-,-)  \\ e_{1R}(-,-)\end{array}\right);
  & \qquad &
 \psi'_{1,R}\equiv 
  \left(\begin{array}{c} \nu'_{1R}(+,-) \\ e'_{1R}(+,-)\end{array}\right);
   \qquad 
   \nonumber \\ [1ex]
 \psi_{2,L}\equiv 
   \left(\begin{array}{c} \nu_{2L}(-,+) \\ e_{2L}(-,-)\end{array}\right);
  &\qquad&
 \psi'_{2,L}\equiv 
    \left(\begin{array}{c} \nu'_{2L}(+,+)\\ e'_{2L}(+,-)\end{array}\right);
  \nonumber \\
 \psi_{2,R}\equiv 
    \left(\begin{array}{c} \nu_{2R}(+,-)\\ e_{2R}(+,+)\end{array}\right);
&   \qquad &
 \psi'_{2,R}\equiv 
  \left(\begin{array}{c} \nu'_{2R}(-,-)\\ e'_{2R}(-,+)\end{array}\right).
\label{leptons}
 \eea

Let us note the mode expansion of  a generic field, $\varphi(x^\mu,y)$ 
with given $Z_2\times Z'_2$ quantum numbers $(z_1,z_2)$:
 \be 
 \varphi^{(z_1,z_2)}(x^\mu,y) = 
 \sum_n^\infty \xi^{(z_1,z_2)}_n(y)~\varphi_n(x^\mu) ; 
 \label{kkexp}
 \ee
for $z_1,z_2 = \pm$ and where the Fourier
parity  eigenfunctions,  $\xi^{(z_1,z_2)}$, properly normalized on the
interval $[0,\pi R/2]$,  are given by 
\bea
\xi^{(+,+)}_n &=&{2 \eta_n\over \sqrt{\pi R}} \cos{(2n) y\over R};
\quad \quad ~~
\xi^{(+,-)}_n = {2\over \sqrt{\pi R}} \cos{(2n-1) y\over R};
\nonumber \\
\xi^{(-,+)}_n &=& {2\over \sqrt{\pi R}} \sin{(2n-1) y\over R};
\quad
\xi^{(-,-)}_n = {2\over \sqrt{\pi R}} \sin{(2n) y\over R};
\label{modes}  
\eea
where $\eta_n$ is $1/\sqrt{2}$ for $n=0$ and 1 for $n>0$. 
Notice from here that only fields with 
$Z_2\times Z'_2$ quantum numbers (+,+) have zero modes. 
For all others $n>0$. 
Thus, we conclude that indeed, after orbifolding
the gauge group is
$SU(2)_1\times U(1)_{Y}\times U(1)_{Y'}$ 
and the zero mode fermion content is
same as the standard model plus an additional neutrino per 
family which is
a sterile neutrino since it does not couple with the $W$ and
$Z$ bosons of the standard model. This is one of the first predictions of
the model. This prediction is purely a consequence of putting the fermions
in the bulk and therefore different from Ref.~\cite{nandi}. As we show
later on in this paper, the sterile neutrino can acquire a large mass and
will therefore decouple from the low energy spectrum. 
From now on we identify  the fermion fields having zero modes  
with the self-explaining standard notation:
$Q, L, u_R, d_R, e_R$ and $\nu_s$. 
Moreover, from here on, we will refer to $SU(2)_1$ as to $SU(2)_L$ and 
consequently to $SU(2)_2$ as  $SU(2)_R$, 
which are perhaps more meaningful to us.

Let us now discuss the Higgs sector of the model. We need Higgs bosons to
break the remaining $U(1)_{Y'}$  and the standard model gauge group, 
as well as to give mass
to the fermions. We choose the minimal set required for the purpose i.e.
a bidoublet $\phi(2,2,0)$ and doublets $\chi_L(2,1,-1)$ and
$\chi_R(1,2,-1)$. We assign the following $Z_2\times Z'_2$ quantum numbers
to the various components of the Higgs bosons:
\be
\phi \equiv 
\left(\begin{array}{cc} \phi^0_u(+,+) & \phi^+_d(+,-)\\
   \phi^-_u(+,+) &  \phi^0_d(+,-)\end{array}\right);\quad
\chi_L\equiv \left(\begin{array}{c} \chi^0_L(-,+) \\
   \chi^-_L(-,+)\end{array}\right); \quad
\chi_R\equiv \left(\begin{array}{c} \chi^0_R(+,+) \\
   \chi^-_R(+,-)\end{array}\right) .
 \ee
They are consistent with the generic $Z_2\times Z'_2$ 
transformation rules for scalar fields. For instance, 
$\chi\rightarrow \pm P \chi$; and  $\chi\rightarrow \pm P' \chi$,
respectively.
We see from this assignments that the only fields that have zero modes 
are $(\phi^0_u, \phi^-_u)$ and $\chi^0_R$. The former
doublet acts like the standard model doublet. 
We will assign vacuum expectation values (vev's) to
$\langle\phi^0_u\rangle = v_{wk}$ and
$\langle\chi^0_R\rangle= v_R$. 
The first vev breaks the standard model symmetry whereas
the second vev breaks the $U(1)_{Y'}$ symmetry, or equivalently, it breaks 
the  group $SU(2)_R\times U(1)_{B-L}$ down to $U(1)_Y$, as in the 
four dimensional theories,
though  in our case there is no $W^{\pm}_R$ zero mode since $SU(2)_R$ is
already broken by the orbifold anyway. 

%%%%%%%%%%%%%%%%%%%%%%
\section{Masses and  mixings}

We now turn our attention to the 
spontaneous symmetry breaking induced by 
$\langle\phi_u\rangle$ and $\langle\chi_R\rangle$.
Let us first note that that only zero mode fields have vevs 
 and therefore, they actually induce 5D mass terms. 
One can then  simplify the analysis of masses
and mixings  just by looking directly at these 5D terms. 
In  the effective 4D theory, the mass induced by the Higgs vacuum will 
generate a global shifting  of the well known Kaluza-Klein (KK) masses,
such that the
actual mass of each level will be given as
 \be 
	m^2_n = m_0^2 + m_{n,KK}^2~;
\label{kkmass}
 \ee
where $m_0$ stands for the 5D Higgs induced mass.
Here $m_{n,KK}$ is the usual KK mass contribution, 
which is given by integer
multiples of  the inverse compactification scale $R^{-1}$.
The values it takes only depend on the parity of the corresponding field, 
so one gets
\be 
m_{n,KK} R = \left\{\ba{ccl}
   2n &\qquad & \mbox{ for $(+,+)$ and  $(-,-)$;}\\
   2n-1&\qquad & \mbox{ for $(+,-)$ and  $(-,+)$} .
\ea\right.
\label{mnkk}
\ee

Similarly as far as mixings go, if they exist among any two fields at the zero
mode level, they will also be global, 
i.e. they will be independent of the KK number.
And there will be no mixings among fields with different KK number.
For instance, on a given 
excited level of SM fields 
there will be the  very same mixings as 
the ones produced at the zero mode level.
Our  model therefore predicts an exact
KK replication of the SM spectrum beyond the compactification scale.
Obviously, 
there will also be towers associated with
all other nonstandard fields that are present on the model. 

\subsection{Charged gauge bosons }

Let us now discuss with some  detail the masses and mixings of the gauge
sector (see also Ref.~\cite{nandi}).
As only one neutral component of the $\phi$ field develops a vacuum, there
is no left-right mixing between the charged gauge bosons. 
One then gets for the KK levels the masses
\bea 
   m_{n,W_L}^2 &=& M_{W_L}^2 + 
    \left({2n\over R}\right)^2 \label{mw1}\\ 
  m_{n,W_R}^2 &=&  {g_R^2 \over 2}\left( v_R^2 + v_{wk}^2 \right) +
  \left({2n -1\over R}\right)^2 ;
\label{mw2}
\eea
for  $M_{W_L}^2=g_L^2 v_{wk}^2/2$ being the mass of the standard $W_L$. 
Here $g_{L,R}$ represent the gauge coupling constants of 
left and right $SU(2)$ gauge groups respectively. 
As $W_R$ does not develop zero modes, its lower
level mass has a nonzero contribution from the compactification scale, as
it is clear from equation (\ref{mw2}) for $n=1$.

\subsection{Neutral gauge bosons}

In order to  simplify the  analysis it is useful to define 
the mixing  angles 
\be 
 \tan\alpha\equiv {g_B\over g_R}; \quad \mbox{ and }\quad 
 \tan\theta \equiv {g_B \cos\alpha\over g_L};
\label{angles}
\ee
where $g_B$ is the coupling constant of $U(1)_{B-L}$. 
Notice that $\tan\theta$ actually corresponds to the 
standard weak mixing angle, in terms of which the massless 
photon field is defined as
\be 
A(x,y) = 
\cos\theta~B_{Y}(x,y) + \sin\theta~W^3_{L}(x,y);
\ee
for $B_Y(x,y) = \cos\alpha~B(x,y)+ \sin\alpha~W^3_{R}(x,y)$ the standard 
hypercharge boson. A Lorentz index must be understood in these equations.
 As it is usual, the orthogonal boson to $B_Y$ is called
$Z'$ and it is given by the combination
$Z'(x,y) = \cos\alpha~W^3_{R}(x,y) - \sin\alpha~B(x,y)$, 
whereas the standard model neutral boson is the zero mode of the field
$Z(x,y) = \cos\theta~W^3_{L}(x,y) - \sin\theta~B_{Y}(x,y)$.

By considering the mass terms induced by the vevs, we notice that 
$\langle\chi_R\rangle$  only  contributes to the  mass of the 
$Z'$ boson, whereas $\langle\phi_u\rangle$ 
generates  masses for both $Z'$ and the standard $Z$, 
and it also introduces a $Z-Z'$ mixing. The photon, as expected, remain
massless. One can now write the $Z-Z'$ mass mixing matrix  as
\be 
 \left(\ba{c c} 
  m_Z^2 & -m_Z^2 \sin\theta\cot\alpha\\
  -m_Z^2 \sin\theta\cot\alpha & m_{Z'}^2
 \ea \right).
\label{zmixing}
\ee
In last equation 
\be 
m_Z^2 = {M_{W_L}^2\over \cos\theta};
\quad \mbox{ and } \quad
m_{Z'}^2 =  \left({g_R^2 v_R^2\over 2 \cos^2\alpha}\right) 
 \left[1 + \left({v_{wk}\over v_R}\right)^2 \cos^2\alpha\right] .
\ee

In the limit when $v_R\gg v_{wk}$ the above mass matrix 
gives a mass correction to the $Z$ boson, 
\be 
M_Z^2 = m_Z^2 - \delta m_Z^2,
\ee
with  $\delta m_Z^2\approx (v_{wk}/v_R)^2~m_Z^2 \cos^4\alpha$.
Correspondingly, one gets $M_{Z'}^2= m_{Z'}^2 + \delta m_Z^2$. 
In the symmetric limit where $g_L=g_R$ the mass correction reads as 
\[{\delta m_Z^2\over m_Z^2}\approx 
{\cos^2 2\theta\over \cos^4\theta}\left({v_{wk}\over v_R}\right)^2.\]
The $Z-Z'$ mixing  is given   by
\be 
\tan\beta\approx 
 \left({v_{wk}\over v_R}\right)^2~ 
{\sin\alpha\cos^3\alpha\over \sin\theta}~
\stackrel{g_L=g_R}{\longrightarrow}~
 \left({v_{wk}\over v_R}\right)^2
{(\cos 2\theta)^{3/2}\over \cos^4\theta}.
\ee

KK modes of  neutral 
gauge fields will have masses which follow the prescription in  
Eq.~(\ref{kkmass}), that is
\be
\left( \ba{c} m_{n,A}^2 \\ m_{n,Z}^2 \\ m_{n,Z'}^2 \ea\right) = 
\left( \ba{c} 0 \\ M_Z^2 \\ M_{Z'}^2 \ea\right) 
+  \left({2n\over R}\right)^2.
\ee
 
%%%%%%%
\subsection{Yukawa couplings and fermion masses}

The most general Yukawa couplings one can write with our matter content,
and which  are invariant under both  gauge and parity transformations are
\be 
h_u \bar Q_1\phi Q_2 + h_d \bar Q_1\tilde\phi Q'_2 + 
h_e\bar \psi_1\tilde \phi \psi_2 +
h'_u \bar Q'_1\phi Q'_2 + h'_d \bar Q'_1\tilde\phi Q_2 + 
h'_e\bar \psi'_1\tilde \phi \psi'_2 +
 h.c.;
\label{Yuk}
\ee
where $\tilde \phi\equiv \tau_2\phi^* \tau_2$ is the charge conjugate
field of $\phi$. 
The matrices $h_{u,d,e}$ and $h'_{u,e}$  are the $3\times 3$
Yukawa coupling matrices in five dimensions. 
The above terms are  invariant under the parity symmetry
that interchanges the subscripts: $1\leftrightarrow 2$ provided
the Yukawa couplings satisfy the following constraints:
\be
h_u~=~h^{\dagger}_u;\quad h'_u~=~h^{'\dagger}_u; \quad
h_e~=~h^{\dagger}_e;\quad h'_e~=~h^{'\dagger}_e; \quad
h_d~=~h^{'\dagger}_d.
\ee
This is the five dimensional realization of the left-right symmetry.
It is worth noticing that there are no trilinear couplings that involve
$\chi$ doublet fields. 
Moreover, there are no couplings in the theory that may give rise to a
Dirac mass for the neutrino. 
This is one of our most interesting results.
From above terms one can read out 
those  couplings that give rise to the standard model fermion masses,
\be
{\cal L}~=~h_u \bar Q\phi_u u_R + h_d \bar Q\tilde\phi_u d_R + 
h_e \bar L\tilde \phi_u e_R + h.c.
\label{yukawa}
\ee
As it is obvious, generation mixings will come through the Yukawa
couplings. Also important to note is the fact that while $h_{u,e}$ are
hermitian matrices, $h_d$ is not; this fact has important implication for
the
nature of the quark mixings involving right handed quarks. To see this we
first note that $h^{diag}_{u, e} = V_{u,e} h_{u,e} V^{\dagger}_{u,e}$
whereas $h^{diag}_{d} ~=~ V_d h_d U^{\dagger}_d$. Thus the $U_{CKM}~=~
V^{\dagger}_uV_d$ whereas the corresponding right handed charged current
mixing matrix for quarks is $U^{R}~=~V^{\dagger}_uU_d$. Thus unlike the
case of standard left-right models, the left and right handed quark
mixings are different from each other.

The `chiral partners' of the standard model fields, i.e. those that come
in the same 5D representation, will have similar couplings, 
$\bar Q_{1R}\phi_u u_{2L} + \bar Q_{1R}\tilde\phi_u d_{2L} + 
\bar \psi_{1R}\tilde \phi_u e_{2L} + h.c.$,
and so, they will get an equal contribution to its mass from the vacuum as
that of its SM partner.
The final mass spectrum will have the SM fermions with usual
masses  at the zero mode level. 
Extra degenerate pairs of (excited) massive  
fermion states will be present above the compactification scale. 
Thus, at each excited level, the spectrum of particles 
will duplicate the SM one. The masses of these KK modes will be of the
form 
\be
 m_n^2 = m_{SM}^2 + \left({2n\over R}\right)^2.
\label{smkk}
\ee

There will also be  some extra particles in the KK spectrum, 
those that get mass contributions from the Yukawa couplings: 
$u'_1$, paired to $u'_2$; $d'_1$ paired to $d_2$ and $e'_1$ paired to 
$e'_2$. Such mass terms will shift the KK mass: $(2n-1)/R$.
The neutral  fermions $\nu_2$, and $\nu_1'$ will all get  Dirac masses:
$(2n-1)/R$. 
Corresponding to  the fields $\nu_1$
and $\nu_2'$ we get  KK modes  with Dirac masses $2n/R$. 
At this point, only the
standard neutrino $\nu_L\equiv \nu_{0,1L}$, and the sterile neutrino,
$\nu_s\equiv\nu'_{0,2L}$,  remain massless, as we already have
anticipated. 
These fields, however, may acquire masses from non renormalizable
operators as we will discuss later on.

\subsection{Scalar sector}

As for the scalar fields. At the zero mode level there will be two
massive neutral scalars. The standard Higgs, $H^0 = \phi_0$ with a mass
of order $v_{wk}$, as usual, and a heavier field,
$\chi_{0R}$, with a mass $\sim v_R$. 
Clearly, they will come accompanied with  
their own tower of excited modes, with masses shifted as indicated in
Eq.~(\ref{smkk}). 
What is perhaps more interesting to notice  is that there will
be no KK modes associated to $\phi^-_u$ nor to $\chi^-_R$, both the fields
(so, both theirs towers)  have been  absorbed for  the $W_{L,R}$ gauge
bosons get masses through the  Higgs mechanism.
$\phi_d$, and $\chi_L$ towers, on the contrary, will be present with mass 
spectra following equations (\ref{kkmass}) and ({\ref{mnkk}), 
according to the parity of each field.

%%%%%%%%%%%%%%%%%%%%%%%%%%%%%
\section{Gauge boson fermion couplings}

Gauge couplings on the theory under consideration
are essentially five dimensional. They are generically
of the form:
\be 
{\cal L} = g_{5D}~ G_M (x,y)~ J^M (x,y);
\label{lint}
\ee
where $J^M =\bar\Psi\gamma^M\Psi$ 
is the (vector-like) five dimensional fermion courrent coupled to a
gauge boson $G$. 
Note that the effective four dimensional gauge coupling constant, $g$, and the 
five dimensional one, $g_{5D}$, are related through the simple scaling: 
$g = \xi^{(+,+)}_0~g_{5D}$.
Notice also that $g_{5D}$ is a dimensionful quantity whose mass dimension
balance the higher dimensionality of the fields, whereas $g$ is dimensionless. 
This scaling
has been already taken into account on all the previous analysis such that
$g_{L,R,B}$ were taken as the actual four dimensional couplings.

Now, in order to do the KK decomposition of the theory one should go to the
unitary gauge where the KK modes of the gauge boson get well defined
masses by absorbing the modes associated to its own fifth Lorentz
component. In such a gauge one takes $G_5 = 0$, which reduces 
the Lorentz index in Eq.~(\ref{lint}) to $\mu$. 
This gauge fixing does not preclude 
the existence of the effective four dimensional gauge invariance associated
to the zero modes. 
On such a gauge we get the general effective couplings among KK modes
\be 
{\cal L}_{eff} \equiv \int dy~{\cal L} = 
\sum_{m n}~ g~\left[{G_{m n,\mu}^L}(x) {J_{mn}^L}^\mu(x)
 +{G_{m n,\mu}^R}(x) {J_{m n}^R}^\mu(x) \right] .
\label{leff}
\ee
Here, the left  and right handed  fermion courrent 
are  given in terms of the excited modes
${J_{mn}^{L,R}}^\mu = \bar\Psi_{m~L,R}~\gamma^\mu~\Psi_{n~L,R}$, whereas
$G_{m n,\mu}^{L, R}$ stands for 
\be 
G_{m n,\mu}^{L,R} (x)\equiv 
\sqrt{\pi R \over 2}~\int dy~ \xi_m^{L,R}(y)~\xi_n^{L,R}(y)~ G_\mu(x,y).
\label{gcoupling}
\ee
The right hand side of last equation can be expanded by introducing the 
KK expansion of $G$. This procedure indicates
that only  excited gauge boson modes that conserve 
the KK number at the vertex contribute to the above couplings. 
Indeed, the explicit  expansion of Eq.~(\ref{gcoupling}) 
involves integrals of the type $\int\xi_n\xi_m\xi_\ell$,
which generically give the result:
$\delta_{\ell,m+n} + \delta_{m,n+\ell} + \delta_{n,\ell+m}$. 
This can also be seen as  a consequence that translational invariance
along the fifth dimension  has
not been explicitly broken in the theory. 
Particularly, at the zero mode level only the $G_{0,\mu}$  couples 
to the purely zero mode courrent, ${J_{00}^{L,R}}^\mu$.
This has the consequence that KK modes could only be produced by
pairs at colliders. 
Following the above prescription, it would be 
enough to write down the couplings at the level of the fifth dimensional
theory to be able to read out those of the effective theory.

\subsection{Charged currents}
As there is no  left right mixing between charged bosons, 
one can write the lagrangian of the charged currents straightforwardly,
\be 
{\cal L}^{cc} = {g_L\over \sqrt{2}} W_{L,\mu}^- J_1^{c,\mu} + 
{g_R\over \sqrt{2}} W_{R,\mu}^- J_2^{c,\mu} + h.c.
\ee
{}For $J_1^{c,\mu} = J_{1L}^{c,\mu} + J_{1R}^{c,\mu}$, with 
\be 
J_{1L}^{c,\mu} = \bar u_L\gamma^\mu d_L + \bar \nu_L \gamma^\mu e_L 
+ \bar u'_{1L}\gamma^\mu d'_{1L} + \bar \nu'_{1L} \gamma^\mu e'_{1L} ;
\ee
that contains the standard charged courrent as its zero mode component.
$J_{1R}^{c,\mu}$, on the other hand, is given by a similar expression with
all fields changed by its chiral partners 
(that is by taking  $u_L\rightarrow u_{1R}$, an so on), that gives
$J_{1R}^{c,\mu} = \bar u_{1R}\gamma^\mu d_{1R} + 
\bar \nu_{1R} \gamma^\mu e_{1R} + 
 \bar u'_{1R}\gamma^\mu d'_{1R} + \bar \nu'_{1R} \gamma^\mu e'_{1R} $.
Last does not contain any zero mode, thus, 
it is absent on the low energy  level.
A CKM matrix acting on the family space should be understood. 
Also note that a normalization
factor for the gauge coupling has been omitted for simplicity.

The charged courrents coupled to $W_R$ are given by 
\be 
J_{2R}^{c,\mu} = \bar u_{R}\gamma^\mu d_{2R} + 
  \bar u'_{2R}\gamma^\mu d_{R}  + 
  \bar \nu_{2R} \gamma^\mu e_R  + 
  \bar \nu'_{2R} \gamma^\mu e'_{2R} ;
  \label{rcourrent}
\ee
That contains the couplings to the right handed SM fields. 
Note, however that
$W_R$ does not couple $u_R$ to $d_R$, but rather  to the $d_{2R}$
field, which belongs to the same representation. 
This has dramatic implications on the phenomenology as we shall mention
below. 
Next,
$J_{2L}^{c,\mu}$ is, again, just the chiral partner of $J_{2R}^{c,\mu}$,
and it reads $J_{2L}^{c,\mu} = \bar u_{2L}\gamma^\mu d_{2L} + 
  \bar u'_{2L}\gamma^\mu d'_{2L}  + 
  \bar \nu_{2L} \gamma^\mu e_{2L}  + 
  \bar \nu_{sL} \gamma^\mu e'_{2L}$;
There is no zero
mode component on these interactions, thus, any correction to SM processes 
due to this terms will appear only through loops, and thus more suppressed
that in the case of four dimensional theories. Furthermore, in general the 
left and the right CKM matrices are not related (similar to the
``non-manifest'' case of standard left-right models). 
 These properties of our model are different from the model presented in
 Ref.~\cite{nandi}.

%%%%%%%%%%%%
\subsection{Neutral currents}

After performing the rotations introduced in the previous section (used to
get gauge mass eigenstates) one gets the following neutral current
interactions:
\be 
{\cal L}_{NC} = 
\left[e A_\mu Q ~+~
\left({g_L\over \cos\theta}\right)~ Z_\mu~ A_{NC}~+~
\left({g_L\over \cos\theta}\right)~ Z'_\mu~ B_{NC} \right]~
J^{NC,~\mu} 
\ee
where the neutral current has the contribution of all 
fermion representations in Eqs.~(\ref{quarks}) and (\ref{leptons}): 
\be 
J^{NC,~\mu}= \sum_{i}\bar Q_i\gamma^\mu Q_i + 
\bar \psi_i\gamma^\mu \psi_i.
\ee
The zero mode components of this current are read out as
\be 
J^{NC,~\mu}_{00,L}= \bar Q\gamma^\mu Q  + \bar L\gamma^\mu L + 
\bar \nu_{sL}\gamma^\mu \nu_{sL} ; 
\qquad 
J^{NC,~\mu}_{00,R}= \bar u_R\gamma^\mu u_R  + \bar d_R\gamma^\mu d_R + 
\bar e_R\gamma^\mu e_R .
\ee
that one identifies as the SM neutral current elements.
The effective couplings, $A_{NC}$ and $B_{NC}$,
follow the general prescriptions:
\[
 A_{NC} = 
\left( \sin\theta \sin\beta \tan\alpha + \cos\beta\right) T_{3L} + 
{\sin\theta \sin\beta \over \cos\alpha\sin\alpha}  T_{3R} - 
(\sin\theta \sin\beta \tan\alpha + \cos\beta\sin^2\theta) Q  ;
\]
and
\[
 B_{NC} = 
 (\sin\theta \cos\beta \tan\alpha - \sin\beta) T_{3L} + 
{\sin\theta \cos\beta \over \cos\alpha\sin\alpha}  T_{3R} - 
(\sin\theta \cos\beta  \tan\alpha  - \sin\beta \sin^2\theta) Q.  
\]
It is illustrative to notice that in the further limit when 
$v_R\rightarrow \infty$, that means that $\sin\beta\rightarrow 0$, one
recovers the standard coupling of the $Z$ boson from $A_{NC}$. 
In the case when one takes  $g_L=g_R$, above expressions reduce to 
\be 
A_{NC} = 
\left(\cos\beta + 
   {\sin\beta \sin^2\theta\over (\cos 2\theta)^{1/2}} \right) T_{3L} +
 {\sin\beta \cos^2\theta\over (\cos 2\theta)^{1/2}}  T_{3R} -
\sin^2\theta \left( \cos\beta + 
   {\sin\beta\over (\cos 2\theta)^{1/2}} \right) Q ;
 \ee
and 
\be
 B_{NC} = 
 \left({\cos\beta\sin^2\theta\over (\cos 2\theta)^{1/2}} + 
   \sin\beta \right) T_{3L} +
 {\cos\beta \cos^2\theta\over (\cos 2\theta)^{1/2}}  T_{3R} -
\sin^2\theta \left( {\cos\beta \over (\cos 2\theta)^{1/2}} - 
   \sin\beta \right) Q .
\ee

%%%%%%%%%%%%%%%%
\section{Neutrino masses without the bulk singlet  neutrino}

Lets us now get back to the problem of neutrino masses. As we mentioned
already, standard and sterile neutrinos do not get Dirac masses from 
Yukawa couplings [see Eq.~(\ref{yukawa})]. 
The physical reason is two fold. First, the zero mode 
right handed neutrino is missed on the theory. Second, the potential
Yukawa terms, $\bar \psi_1\phi \psi_2$ or $\bar\psi_1\phi\psi'_2$, 
that  may give rise to a large 
Dirac neutrino mass, and 
that one could expect  from the matter
content in Eq.~({\ref{leptons}), are not
invariant under the parity symmetries, $Z_2\times Z'_2$. 
Thus the neutrino is massless at the renormalizable level of
the effective four dimensional theory. 
This is a completely new feature for this class of theories with left
right symmetry and completely distinct from the four dimensional
left-right models.
The neutrino mass does not arise in the present model from see-saw
mechanism, and
thus, there is not immediate constraint on $v_R$ scale from this sector.
This allows for the possibility of a sufficiently small $v_R$ so as to be
accessible to the  next generation collider experiments.

On the other hand, one common problem of theories with a low fundamental
scale is a potentially dangerous   large neutrino mass that comes  from
non renormalizable operators, such as $(LH)^2$. 
Question is then whether 
such a problem may be also present on our theory. 
As we now show, this is not longer a problem.  Clearly such an operator,
coming from its higher dimensional relative: $(\psi_1 \phi)^2$,  
is forbidden by the conservation of $B-L$ on our model. Moreover, though
the coupling $(\psi_1\chi_L)^2$ is perfectly possible, there is no neutrino
mass induced from this operator due to the lack of a vev for $\chi_L$. 
Therefore the model is completely safe from dangerous operators
that could give large neutrino masses.

To see how neutrino mass arises in this theory, let us note that
there are three classes of  non-renormalizable operators of higher
dimensions that remain invariant under all the symmetries of the theory,
and
which  contribute to neutrino masses. 

(i) There are operators connecting
the active left handed neutrinos to themselves: i.e.
$O_1 \equiv \psi_1^TC_5 \psi_1 \phi\phi \chi_R\chi_R/M^5_*$, where 
$C_5\equiv\gamma^0\gamma^2\gamma^5$, where we have omitted the family
index. 
Notice that it has dimension 10 on 5D.
It generates, at the four
dimensional theory, the effective couplings~\cite{carlos}
\be 
{h\over (M_*R)^2}~ {(L\phi_u\chi_R)^2\over M_*^3  };
\ee
with $h$ the dimensionless coupling. 
This operator induces a sufficiently
small Majorana neutrino mass, 
\be 
\label{numass}
m_\nu = {h~v_{wk}^2 v_R^2 \over ({M_*}R)^2~ M_*^3 } \approx ~h\cdot~ 1 ~eV .
\ee
where the right hand side has been estimated using 
$v_R\approx 1/R\approx 1~TeV$ and $M_*\approx 100$ TeV. 
A soft hierarchy in the couplings (say $h\sim 0.01$) should provide the
right spectrum on neutrino masses. 

(ii) The second class of operators connect $\nu$ to $\nu_s$ and have the
form in lowest order 
$O_2 \equiv \psi^T_1\tau_2\phi \chi_R
\chi^T_R \tau_2 C_5\psi'_2/M^{7/2}_*$. This operator after
compactification has a magnitude 
$\simeq \frac{v_{wk} v^2_R}{M^2_*(M_*R)^{3/2}}\simeq 10$ keV 

(iii) The last class connects the left handed neutrinos that transform
under the $SU(2)_R$ group to themselves and have the form
 $O_3\equiv (\psi'_2\chi_R)^2/M^2_*$. They contribute to the
$\nu_s$-$\nu_s$ entry and have magnitude after compactification
estimated to be $\simeq \frac{v^2_R}{M^2_*R}\simeq 1-10$ GeV. The full
$6\times 6$ $\nu-\nu_s$ mass matrix
has a seesaw like form
and on diagonalization, leads to an effective mass for the light neutrino in
the range of $0.1$ eV or so.

One should note that, in the present model, there is no need to invoke 
extra bulk neutrinos living in a larger number of extra
dimension~\cite{nima2} to get small neutrino masses. This is a major
advantage of this model over the original models where a bulk neutrino was
needed to get neutrino masses.

 There should be, nevertheless, at least two 
more flat extra large dimensions to compensate for 
the gap between the fundamental and Planck scales. 
The size of such dimensions would be of order micrometers. 
Current bounds on $M_*$ coming from bulk graviton effects~\cite{exp1}
will apply. A single warped extra dimension~\cite{rs}, instead, 
could also provide a  good explanation for the smallness of $M_*$.

\subsection{sterile neutrino mass and cosmology}
As noted in the previous subsection, there is a heavy mass ($\sim $ 1-10
GeV) sterile neutrino in this model due to the presence of the operator
${\psi'_2}^T C_5\psi'_2 \chi_R\chi_R$, that induces a Majorana neutrino
mass,
\be 
 m_{\nu_s}^2 = h_s {v_R^2\over M_*^2 R};
\ee
We do not enter into the
detailed cosmological implication of such heavy sterile neutrinos except
to note that they can annihilate via the exchange of $Z'$ boson into
lighter quarks and leptons. For a 1-10 GeV sterile neutrino annihilating
via the exchange of 1 TeV $Z'$, the typical temperature at which the
$\nu_s$ goes out of equilibrium is around $T_*\sim M_{\nu_s}/15$. From
this we estimate that the contribution of $\nu_s$ to the total energy
density of the universe at the BBN epoch i.e. $T\simeq 1$ MeV is
equivalent about a tenth of a neutrino. Therefore this does not effect
the usual Helium synthesis scenarios of the standard big bang model.

%%%%%%%%%%%%%%%%%%%%%%%%%

\section{Phenomenology}

There is a series of dramatic implications on the phenomenology of the
present model that makes it completely different from all previous
left-right constructions: 

\begin{itemize} \item[--] 
There are no tree level contributions to muon decay from the new particles
of the theory. This is due to the conservation of the KK number on the
vertex, which forbids the mediation of the decay by any KK gauge bosons. 
That includes the $W_R$ boson which itself is a KK mode (the
lightest one in its tower), and all KK modes of $W_L$.  
Also, there are no new extra channels 
for the process  since all  lighter particles are only the ones in the 
standard model.  
Thus, no good bounds on the masses of the new particles in the model can
be obtained from here.

\item[--]
For similar reasons, $W_R$ has no relevant contributions to 
neutrinoless double beta decay. At tree level, all external legs on the
diagram are zero mode (SM) particles. This constrains the internal
particles to be zero modes too. The process will still 
take place, but only  as it is usually expected from the 
fact that neutrinos arise as Majorana particles in the
model (see the previous section). 

\item[--]
More generally, due to the conservation of the KK number, 
the excited modes can only contribute to SM processes through loops. 
Thus, most of their contributions would be very suppressed, since all
internal lines in the loop will get a heavy particle propagator on it.  

\item[--] Conservation of fifth momentum implies that the lightest mode
with a non zero KK number will be stable. This may, in principle, be
either hadronic or leptonic depending on relative Yukawa couplings.
%%%%%%%%%%%%%
From our analysis of the  mass spectrum (in section iii) 
one can see that the lightest KK particle (LKP)
is likely to be a KK right handed neutrino, the one out of $\psi'_1$. 
Reason is twofold: Being parity odd (even) under $Z_2'$ ($Z_2$)
it has  the lower KK  mass possible: $~1/R$. 
It also gets the smaller contribution to its mass 
from the vacuum. Last comes from similar non renormalizable operators  
to those  responsible for the standard neutrino mass, which  actually
give a negligible contribution. 
A barionic candidate seems unlikely. All KK quarks 
get vacuum mass contributions from the Yukawa couplings in
Eq.~(\ref{Yuk}), which are expected to be of the 
same order than the SM ones.
Note also that the  lightest KK photon has a mass $2/R$. 
%%%%%%%%%%%%%

\item[--] 
In four dimensional versions of the model (see also Ref.~\cite{nandi}),
the stringent bound on $W_R$ mass usually comes from the contributions of
$W_R$ to the $K-\bar K$ mixing~\cite{soni}.  However, in
the present case the situation changes considerably. The usual 
$W_L-W_R$ box graph that gives the largest contribution to this process
in the standard left-right models with equal left and right quark
mixings does not exist since the $W_R$ coupling involves KK modes as can
be seen from Eq.~(\ref{rcourrent}) and does not couple 
to the familiar charged current $\bar u_R\gamma^\mu d_R$ where $u,d$ are
both light quarks (or zero modes in this model).
There would of course be a diagram similar to the Standard
model one, with $W_R$ running on both internal boson lines instead of
$W_L$. In this case the internal fermion lines will also be the KK modes
of the fermions of the model. This
will give new nonzero contribution to the $K-\bar{K}$ mixing
 from right handed currents. 
Now, since all KK modes can run in the loop, their contribution
should be
summed up in the total amplitude. Notice that all internal lines on these
last diagrams will correspond to KK exited modes of the same level.
 Thus, they will come with a large suppression
due to the heavy masses in the propagators. These contributions are
suppressed compared to the left-handed ones by a factor of
$\left(\frac{M_{W_L}}{M_{W_R}}\right)^4$. The limits on the $W_R$ mass
from these considerations are therefore very weak.

\item[--] 
In the four dimensional left-right models, there are flavor changing
neutral currents mediated by the second standard model Higgs doublet in
the bidoublet. This in effect pushes the limit on the right handed scale 
to 6-8 TeV\cite{ecker}. In our model however, since the second SM doublet
has (+-)
$Z_2\times Z'_2$ symmetry, it couples only to KK modes to the SM fermions
and therefore does not lead to such effects. Also constraints coming
from processes such as $b\rightarrow s +\gamma$\cite{rai} are also absent
in this model. 

\item[--]
Limits on R may come from collider physics from the production of the KK
modes of the $W_L$ boson\cite{dob} and are in the range of 400 to 800 GeV. 
The important point to note is that the $W_R$ boson as well as all KK
modes are produced in pairs both in $e^+e^-$ as well as hadron colliders.
The decay modes of the $W_R$ are: $eN$, $d\bar{U}$ etc where $N,U$ are the
KK modes.

\item[--]
A limit on $v_R$ comes from the $Z-Z'$ mixing.
The analysis of this is as in the standard four dimensional theories
due to the universality of the mixing~\cite{nandi}. 
One gets $v_R\gs 800$ Gev. This limit is likely to go up once LHC is
turned on to the TeV range\cite{ferrari}.

\end{itemize}

\section{Comments and outlook}

\subsection{Baryon non-conservation and six dimensional extensions}
In this model as in other 5-dimensional  brane-bulk models, one can write
a baryon non-conserving operator $Q_1Q_1Q_1\psi_1/M^3_*$, which for low
$M_*$ leads to short life times for the proton. There is also another
baryon nonconserving operator of the form
$Q_1Q_1Q_1\phi Q_2Q'_2Q'_2\chi_R\chi_R/M_*^{23/2}$. This leads to
neutron-anti-neutron oscillation after compactification to four
dimensions. 

To avoid the problem of proton decay, one may proceed
in one of the two following ways: (i) consider a fat brane with quarks
located at a separate point from the leptons~\cite{schmaltz} or (ii)
embed the model into a six dimensional space 
time~\cite{dobrescu,Gouverneur,ponton}. The
second alternative  is attractive for our point of view since it
automatically brings in the right handed neutrino into picture and
suggests a left-right symmetric model. 

To see briefly the consequences of a six dimensional embedding, note that
the  six dimensional embedding comes at a minimal 
cost with the same fermion sector as in
Eqs.~(\ref{quarks}) and (\ref{leptons}).
Our five dimensional fermion representations 
can be straightforwardly written in terms of six dimensional ones. Six
dimensional fermions are eight-component fields, but they may have a
well defined chirality
through the operator $1\pm \Gamma^7$, 
with $\Gamma^7$ being the product of eight by eight Dirac matrices:
$\Gamma^7 = \Gamma^0\Gamma^1\cdots \Gamma^5$. After chiral projection, they
reduce to four component spinors as required in the five dimensional
theory. 
The irreducible gauge  
and gravitational anomalies naturally cancel for the 
six dimensional chirality assignments:
\be 
{\cal Q}_{1~+}~,~ {\cal Q}'_{1~+}~,~ 
{\cal\psi}_{1~+}~,~{\cal\psi}'_{1~+}~,~
{\cal Q}_{2~-}~,~ {\cal Q}'_{2~-}~,~ 
{\cal\psi}_{2~-}~,~{\cal\psi}'_{2~-}~,~
\ee

One of the first implications of such an embedding is that
all dangerous proton decay inducing operators are naturally forbidden
~\cite{ponton}.
This happens due to the extended Lorentz  symmetry of the theory 
which is not broken in the Lagrangian, and which contains some discrete
subgroups that only allow proton decay through operators of dimension 
15 or higher. Notice that, up to a duplication of the spectrum, the
particle content is the same considered in the analysis of 
Ref.~\cite{ponton}. Thus, the details of the argument will follow same
lines.

Next, six dimensional models 
have the potential of explaining the number of 
generations~\cite{dobrescu,Gouverneur}. 
The argument is based on the fact that the cancellation of global $SU(2)$
anomalies imposes that~\cite{vafa}
\be 
N(2_+) - N(2_-) = 0~{\rm mod}~ 6;
\label{n2diff}
\ee
where $N(2_\pm)$ is the number of doublets with chirality $\pm$.
This applies to both $SU(2)$ groups.
The relation (\ref{n2diff}) is (non trivially) satisfied only for at 
least three generations.

In the six dimensional version of our model, neutrino mass arises in an
interesting manner. The 
simple neutrino mass operator involving the active neutrinos i.e. $O_1$
alone discussed in sec. v is now forbidden 
 due to the $U(1)_{45}$ symmetry as
is the operator $O_2$ involving only the $\nu_s$. However the $U_{45}$
symmetry allows the operator $O_2$ connecting $\nu$ and $\nu_s$.
Furthermore in six dimensions, $O_2$ has dimension eleven so that after
compactification, it leads to the neutrino mass expression $m_{\nu}\sim
\frac{v_{wk}v^2_R}{M^2_* (M_*R)^3}$. For $M_*\sim 100$ TeV and $R^{-1}\sim
1$
TeV, this gives an $m_\nu \sim 1$ eV for a coupling strength multiplying
these operators of order $0.1$. Again, in this case, we do not need a
bulk neutrino. The
neutrino in this case is however a Dirac neutrino and will therefore
not allow the neutrinoless double beta decay process.

 We will pursue the details of
the six dimensional extension of the left-right model in a separate
publication. 

\subsection{Supersymmetry}

Another interesting alternative to extend the present model is to 
include supersymmetry at the 5D level. Notice that the above described
six dimensional theory, however, can not be trivially supersymmetrized, since
the addition of new representations, as susy scalar and gauge  partners,
may  destroy the properties that led to proton stability and the
potential explanation to number of generations.

The cost of introducing susy on our 5D model 
is minimal for the scalar content  
which now should be duplicated. The effective zero mode theory will be 
just the MSSM plus heavy sterile neutrinos.
However, since one can not write trilinear
couplings in a N=1 5D theory, supersymmetry has to be broken by the
boundary conditions down to the effective N=1 4D susy. This can be easily
done at the fixed point $y=0$ without affecting the other properties of the
model. As a consequence, all Yukawa couplings, Eq.~(\ref{Yuk}), as well as
the Higgs vevs, should now be localized on the border where susy is broken.
The localization of these terms will change the profile of the wave 
function along the fifth
dimension, that will now try to match the presence of a point like source
that comes as a localized mass  term. This can also be understood in terms
of the former Fourier expansion in Eq.~(\ref{kkexp}) by noticing that a 
localized interaction term, as  
$\bar Q_1(x,y) \phi(x,y) Q_2(x,y)~\delta(y)$ for
instance, introduces a global mixing among all KK modes, and thus the basis
has to be rotated to get the proper mass eigenstates. 
In such a case, most of the mentioned phenomenological properties associated
to the universality of mixings and the conservation of KK number fail, so
their implications  are likely to change.

\subsection{Restoration of parity at high energies}

In conventional left-right models, it is well known that as one goes to
extreme high energies i.e. when $Q^2$ in a process is much higher than the
mass square of the $W_R$ and $Z'$ bosons, weak interaction processes
involving left and right handed helicities become equal upto small
corrections of order $m^2_{W_R}/ Q^2$. Similarly, in the early universe
when the temperature $T\gg m_{W_R}$, phase transition takes place
leading to a symmetric phase of the theory and parity is fully
restored. By the same token, as the Universe cools below the parity
restoration scale ($v_R$ scale), symmetry is broken and there are domains
of even and odd parity separated by domain walls. The apriori possibility
that such domain walls can cause extreme anisotropy in the universe always
is a matter of concern for the cosmology of such models. A common way to deal
with such issues is to ``inflate'' the domain walls away and keep the
$v_R$ scale above the reheating temperature so that the walls are not
regenerated again\cite{cline}.

 For conceptual clarity let us explain how approximate
parity restoration occurs in theories with orbifold compactification.
For simplicity, consider the weak interaction contribution to the cross
section for two scattering processes 
$\sigma_{e^-{_L}N}$ and $\sigma_{e^-_RN}$. For energies $E\leq R^{-1}$,
$\sigma_{e^-_R N}=0$ whereas $\sigma_{e-_L}$ is nonzero. As $E\geq
R^{-1}$, $\sigma_{e^-_R N}\neq 0$ and also $\sigma_{e^-_ N}$ receives an
additional contribution both coming from the first KK excitation of
fermions and gauge bosons. The additional contribution to $\sigma_{e^-_L
N}$ and the value of $\sigma_{e^-_R N}$ are nearly equal apart from some 
propagator corrections. As $E\gg R^{-1}$, more and more states contribute
to both processes and the zero mode contribution to $\sigma_{e^-_L N}$
which was nonzero in the beginning starts becoming a small part of the
whole contribution and one has $\sigma_{e^-_ N}\simeq \sigma_{e^-_R N}$.
This is the sign of parity restoration in the class of models we are
discussing. As $E\geq M_*$, then stringy contributions dominate and
presumably both cross sections become equal obliterating the effect of the
orbifold boundary conditions.

 \section{Conclusion}

   In this paper we have presented a five dimensional left-right symmetric
model where the gauge bosons as well as the fermions reside in the 
full five dimensions. The string scale in this model can be as low as a
100 TeV and the radius of the fifth dimension of order of a (TeV)$^{-1}$. 
The gauge symmetry is partially broken by the
orbifold boundary conditions. We show that one can consistently
remove the right handed neutrinos from the zero mode spectrum, 
which contains the standard model. 
The scale of parity breaking could therefore be under a TeV. The orbifold
breaking enables this low scale to be compatible with all known low energy
data. This
model has a number of other features which are different from the standard
four dimensional left-right models. For instance, in our model, the $W_R$
bosons
are produced only in pairs. Therefore hadron colliders have no special
advantage over the $e^+e^-$ collider for testing this model. It predicts
a stable  lepton with mass $R^{-1}$. Furthermore,
the structure of our model
allows a new way to understand the small neutrino masses in low scale
string models without introducing bulk neutrinos.
Due to the completely new structure of this class of models compared to
the standard four dimensional left-right model, there will be many new
phenomenological implications. We have briefly mentioned some of them.
More details of these implications are under consideration.

\section*{Acknowledgements}
APL whishes to thank the Particle Theory group of the University of Maryland
for the warm hospitality during the realization of this work.  
This work is supported by the National Science
Foundation Grant No. PHY-0099544. We are grateful for the partial
research support from the University of Maryland's Center for String and
Particle Theory that enabled this investigation.

\end{document}